# Increasing Randomness Using Permutations on Blocks

Sindhu Chitikela and Subhash Kak

**Abstract:** This paper investigates the effect of permutations on blocks of a prime reciprocal sequence on its randomness. A relationship between the number of permutations used and the improvement of performance is presented. This can be used as a method for increasing the cryptographic strength of pseudorandom sequences.

**Introduction**

Pseudorandom sequences that are algorithmically produced have limited cryptographic applications because the eavesdropper can readily generate them. The complexity of the generation process and the lack of correlation amongst the bits (or digits) of the sequence are important in determining the usefulness of a pseudorandom sequence. A quantum mechanical process can be used to generate a true random sequence but the problem with such an approach is that such sequences cannot be replicated. Classical random sequences also find use in quantum cryptography applications since the random base choices or rotations there, either in the BB84 protocol [1] or the three-stage protocol [2]-[4], must be generated by an algorithmic process.

To develop a method of improving the quality of pseudorandom sequences, the question of a metric for the degree of randomness must be addressed. There are several ways the randomness of a binary sequence is defined statistically [5] and from a computational complexity point of view [6]. The problem of randomness is complicated by entanglement in quantum systems [7],[8] and it shall not be considered here. One popular method of defining randomness of an *n*-bit long sequence *a(i)* is given by the following formula

$$R(sequence) = 1 - \frac{1}{n-1}\sum_{k=1}^{n-1}(|c(k)|)$$

where *c(k)* is the autocorrelation function $c(k) = 1/n \sum_{j=1}^{n}(a_j a_{j+k})$, where the sequence is represented in terms of +1s and -1s. . This is intuitively satisfactory since for a completely random binary sequence this randomness measure is equal to 1 and for a constant sequence the randomness measure is 0. For a maximul length shift-register sequence of period $2^k$ [9], the randomness measure is *1-1/n*. For good pseudo-random sequences, the randomness measure will be a number just less than 1.

Prime reciprocal sequences or d-sequences [10]-[14] have many applications andany pseudo-random sequence can be mapped to a suitable d-sequence. As seen in Figure 1, the randomness measure gets closer to 1 as the period of the d-sequence increases which is perfectly consistent with the theorem that prime reciprocal sequences are normal sequences.

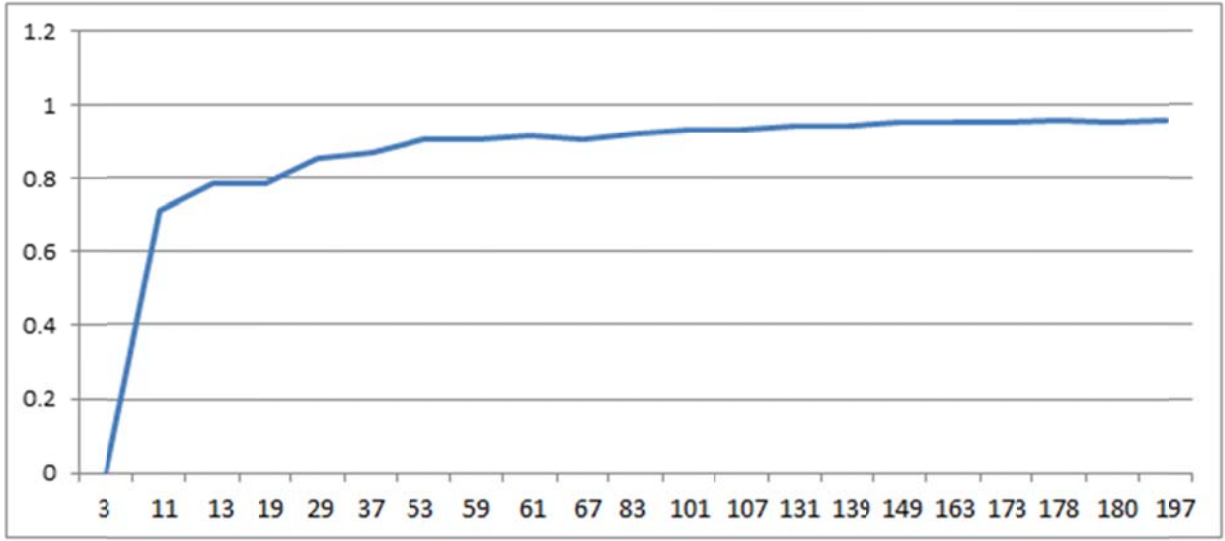

Figure 1. Randomness measure of prime reciprocal sequences to 200

A number $x$ is simply normal in base $r$ if in the decimal of $x$ each of the $r$ possible digits occur with a frequency $1/r$, i.e., $\lim_{n \to \infty} \frac{n_b}{n} = \frac{1}{r}$ for all $b$, where the digit $b$ occurs $n_b$ times in the first $n$ places and a number $x$ is normal in base $r$ if all of the numbers $x, rx, r^2x, ...$ are simply normal in all of bases $r, r^2, r^3, ...$ It follows that when $x$ is expressed as a decimal in the scale of r, every combination $b_1, b_2, b_3, ...$ of digits occurs with the proper frequencies. Thus, the property that a number is normal in base **r** may be reiterated by saying that all the digits $0 - (r - 1)$ occur with equal probability, and that each digit of the sequence is independent of every other digit. Almost all numbers are normal in any base.

Nevertheless, from a practical point of view, given prime reciprocal sequences are not entirely satisfactory. To see this first note that the prime reciprocal sequence $a(i), i = 1,2,3,...$ for prime $p$ (that is the sequence $1/p$ in base 2) can be generated as $a(i) = 2^i \mod p \mod 2$ (Reference [12]):

    *b(0)= 1*
    *b(i+1) = 2b(i)* mod *p*
    *a(i)=b(i)* mod *2*

Maximum length (with period *p-1*) prime reciprocal sequences are generated when *2* is primitive root of *p*.

Although maximum length binary prime reciprocal sequences have excellent autocorrelation properties they have the negative peak of -1 for half the period that reflects the fact that the sequence after half the period is a complementary image of the first half. As example, the binary d-sequence for 1/13 is 000100111011 where the last 6 bits are complements of the first 6 bits.

This means that although the randomness measure of such sequences is high, it is not very useful in this context.

We suggest performing another transformation on the given sequence. In contrast to an earlier preliminary study [15] where groups of bits were mapped to a single bit based on plurality of 0s or 1s to improve autocorrelation properties, here we consider the effect of block permutations on autocorrelation. A number of different random permutations are applied to the blocks of the candidate pseudorandom sequence. We will show that doing so improves the autocorrelation performance considerably. The specific questions that are answered in this paper include a relationship between number of different permutations used and the improvement of performance.

**Choosing Blocks for Permutations**

A d-sequence can be divided into either even number of blocks or odd number of blocks. The performance of the permutation for the d-sequences does depend on whether the number of blocks is even or odd. For example, the d-sequence of the prime number 1277 can be divided into blocks in a variety of ways as 1276 has factors 2, 4, 11, and 29. Here we will consider the division of 1276 into 58 blocks of size 22 bits or 319 blocks of size 4 bits.

In the general case, the sequence $S$ can be represented as the concatenation of blocks $S_1 S_2 S_3 S_4 ...$ We represent an n-permutation by the operator $P_n = P_1 P_2 P_3...$ so that the permutations $P_1$, $P_2$, $P_3$,… are applied in sequence. For example, 3-permutation $P_3$ will work as follows:

$$P_3(S) = P_1(S_1)\ P_2(S_2)\ P_3(S_3)\ P_1(S_4)\ P_2(S_5)\ ...$$

In the first experiment, we consider the d-sequence of length 1276 which is divided into 58 blocks that is $S_1$, $S_2$, $S_3$… $S_{58}$. We generated a random permutation, P, of size 22. This permutation, P is applied on all the 58 blocks of the d-sequence. If the position of each digit is represented with the help of an alphabet as follows.

1 0 1 0 1 0 0 1 1 0 1 1 0 1 1 1 1 0 1 1 1 1

a b c d e f g h i j k l m n o p q r s t u v

P is the permutation "hajblcfedgikovusrqnpmt" and it transforms the given block to 1100110100111111011101. This random permutation "hajblcfedgikovusrqnpmt" is applied on each of 58 blocks of the sequence. We have conducted this experiment many times where the permutation P varies in each experiment. The average of all auto-correlation values is plotted in the graph shown in Figure 2.

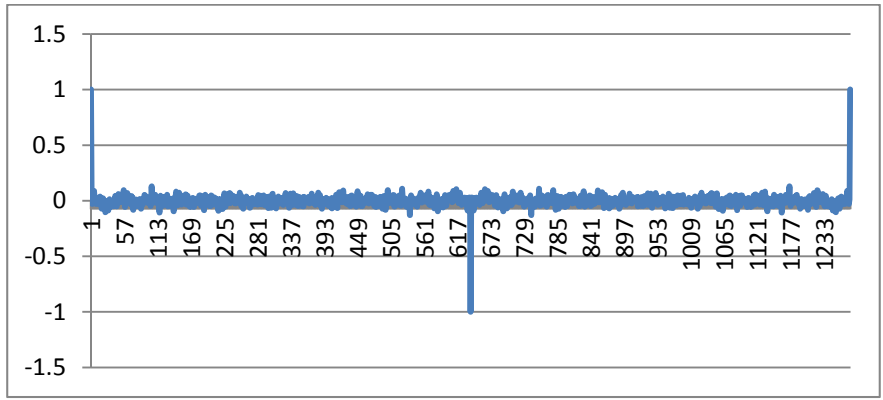

Figure 2 Autocorrelation of the d-sequence with a single permutation applied on its 58 blocks of size 22 digits each

To stress the difference with odd number of blocks, we next consider 319 blocks of size 4 digits each of the d-sequence of 1277. We applied a single permutation P, on all the 319 blocks as we did in the case of even number of blocks. The graph in Figure 3 shows the auto-correlation values of the d-sequence for odd number of blocks. As the autocorrelation function for half the period is less than what it was for the case of even number of blocks, this clearly shows that the performance of permutation process varies for even and odd number of blocks.

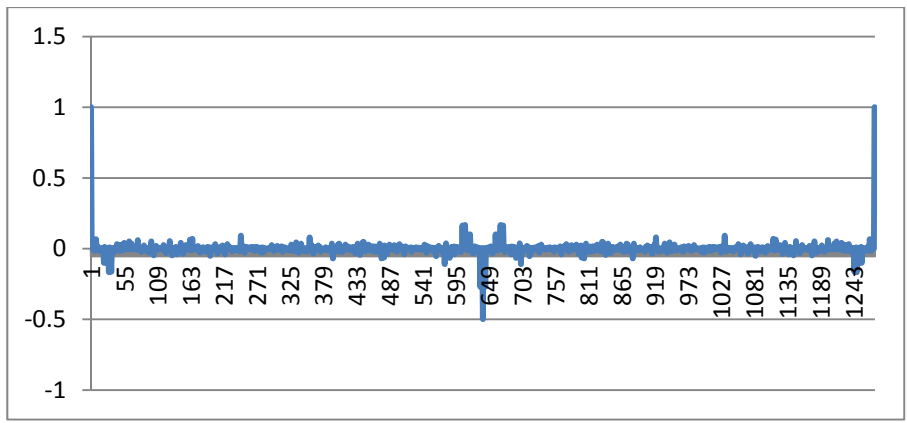

Figure 3 Autocorrelation of the d-sequence with a single permutation applied on its 319 blocks of size 4 digits each Next, as a continuation of the first experiment on the d-sequence for even number of blocks, we generated two random permutations $P_1$, $P_2$ of length 22 each. The permutation $P_1$ is applied on block1 and the permutation $P_2$ is applied on block2. Then the same two permutations $P_1$ and $P_2$ are applied on block3 and block4 respectively. This is repeated for all the 58 blocks of the d-sequence. We conducted the experiment many times where the permutations P1 and P2 are different every time and plotted the average of the autocorrelation values in the graph shown in Figure 4.

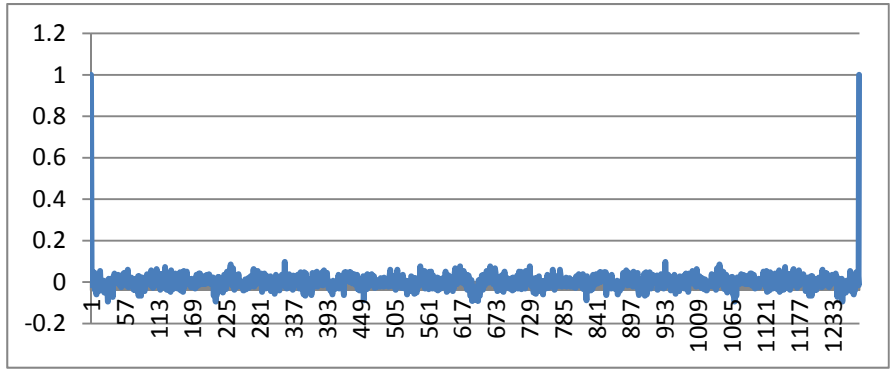

Figure 4 Autocorrelation of the d-sequence of 1277 with two different permutations on 58 blocks of size 22 digits each

Next we consider four random permutations $P_1$, $P_2$, $P_3$ and $P_4$. We applied the permutations P1, P2, P3 and P4 on block1, block2, block3 and block4 of the d-sequence of period 1276. Then, we applied the same four permutations, P1, P2, P3 and P4 on block5, block6, block7 and block8 respectively and this process was repeated till the end of the 58 blocks.

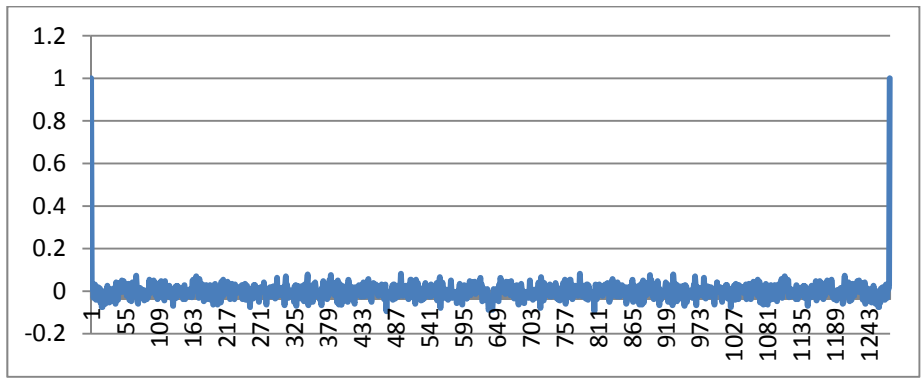

Figure 5 Autocorrelation of the d-sequence of 1277 with four different permutations on its 58 blocks of size 22 digits each

Similarly we considered five, six, seven, eight, nine and ten different permutations on the 58 blocks of the d-sequence of 1277. As a final step, we generated 58 random permutations $P_1$, $P_2 \ldots P_{58}$ on block1, block2…block58 respectively. We conducted the experiment many times where the permutations are different every time and plotted the average of the autocorrelation values in the graph shown in Figure 6.

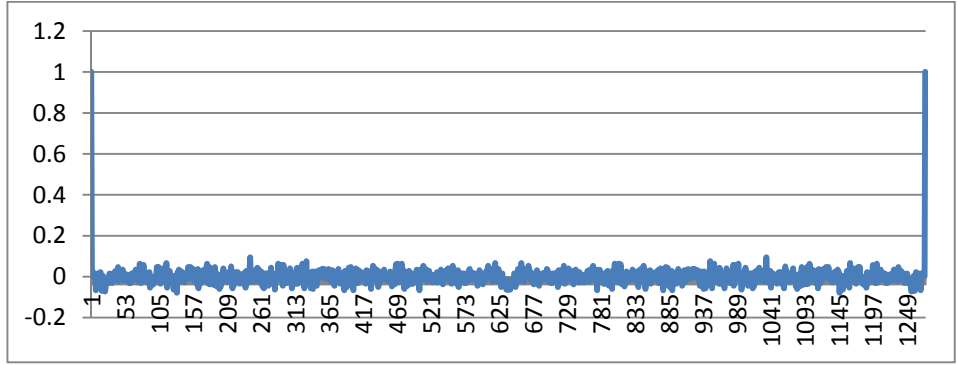

Figure 6 Autocorrelation of the d-sequence of 1277 with 58 different permutations on its 58 blocks of size 22 digits each

**Off Peak autocorrelation for different number of permutations performed on the d-sequence of 1277**

The following table, Table 1 represents the maximum auto-correlation values of the d-sequence of the prime number, 1277. These are the results observed when the above experiments of different permutations are performed on the d-sequence of 1277 which is divided into 58 blocks of size 22 digits each. Table 2 represents the maximum auto-correlation values of the d-sequence of the prime number 1277 for odd number of blocks that is 319 blocks of size 4 digits each.

**Table 1** Absolute maximum of the autocorrelation values of the d-sequence of 1277 which is divided into 58 blocks of size 22 digits each that is even number of blocks

| Number of different permutations | 0 | 1 | 2 | 3 | 4 | 5 | 6 | 7 | 8 | 9 | 10 | 58 |
|---|---|---|---|---|---|---|---|---|---|---|---|---|
| Maximum auto-correlation Value | 1.0 | 1.0 | 0.10 | 0.09 | 0.10 | 0.10 | 0.09 | 0.10 | 0.24 | 0.10 | 0.13 | 0.08 |

**Table 2** Absolute maximum of the autocorrelation values of the d-sequence of 1277 which is divided into 319 blocks of size 4 digits each that is odd number of blocks

| Number of different permutations | 0 | 1 | 2 | 3 | 4 | 5 | 6 | 7 | 8 | 9 | 10 | 319 |
|---|---|---|---|---|---|---|---|---|---|---|---|---|
| Maximum auto-correlation Value | 1.0 | 0.47 | 0.38 | 0.41 | 0.24 | 0.64 | 0.31 | 0.32 | 0.37 | 0.26 | 0.34 | 0.19 |

The striking difference between the two Tables if for the value at 1-permutation where for obvious reasons it makes for no improvement if the number of blocks is even. Also if the size of the blocks is small, the reduction in the value of the off-peak autocorrelation is small.

**Improvement Factor**

The Improvement Factor in the off-peak autocorrelation function of any d-sequence may be measured by the following formula.

*Improvement Factor, I = 1/maximum ( |c(k| ), k ≠ 0*

We considered the improvement factor as a measure of randomness in our experiments. Figures 7 and 8 show the improvement factor for the d-sequence of prime 1277 for different number of permutations.

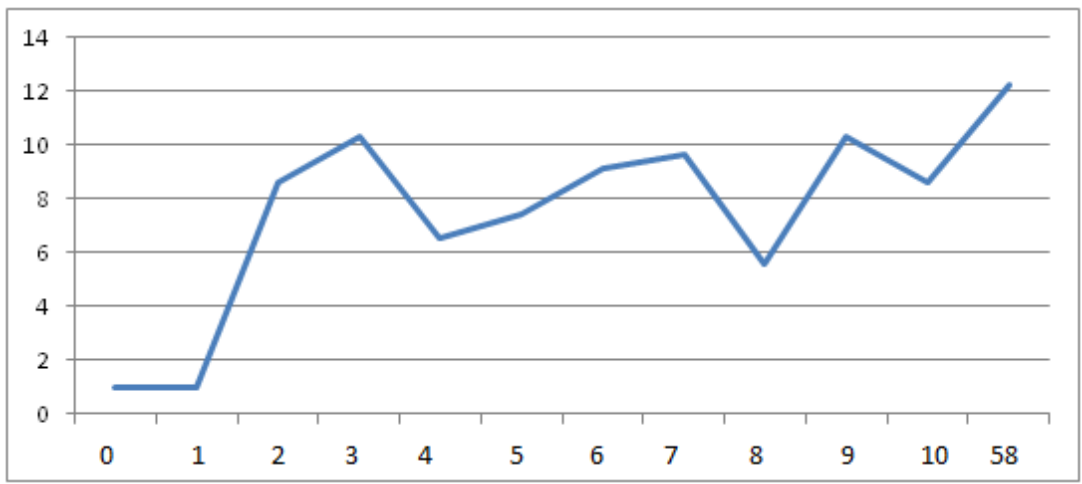

Figure 7 Improvement Factor of the d-Sequence of 1277 when divided into 58 blocks of size 22 digits each that is even number of blocks

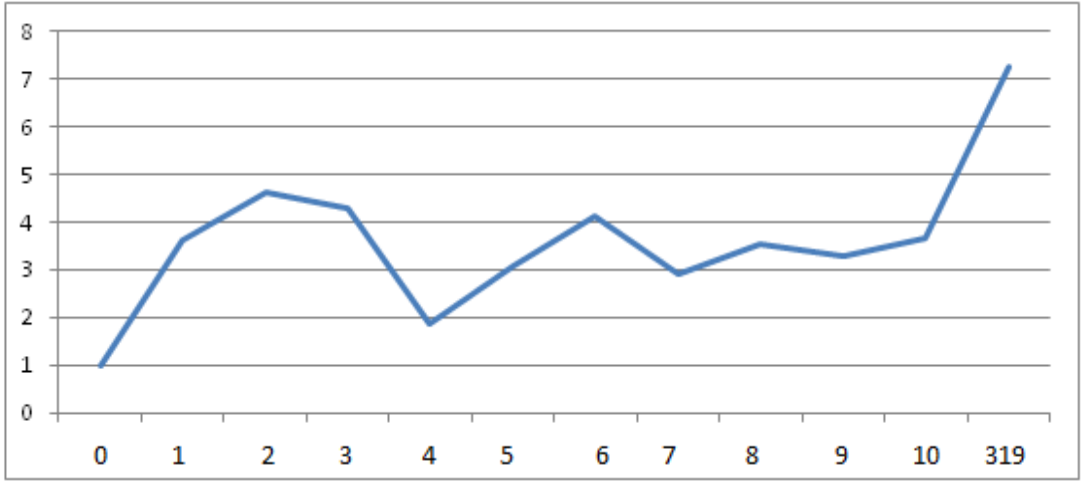

Figure 8 Improvement Factor of the d-Sequence of 1277 when divided into 319 blocks of size 4 digits each that is odd number of blocks

We conducted the above experiments for a large number of primes that lead to maximum length d-sequences. Figures 9 and 10 show the improvement factor of the permuted d-sequence of 1787. Figure 9 shows the improvement factor of the d-sequence of 1787 where it is divided into even number of blocks that is 94 blocks of size 19 digits each. Figure 10 shows the improvement factor of the d-sequence of 1787 where it is divided into odd number of blocks that is 47 blocks of size 38 binary digits each.

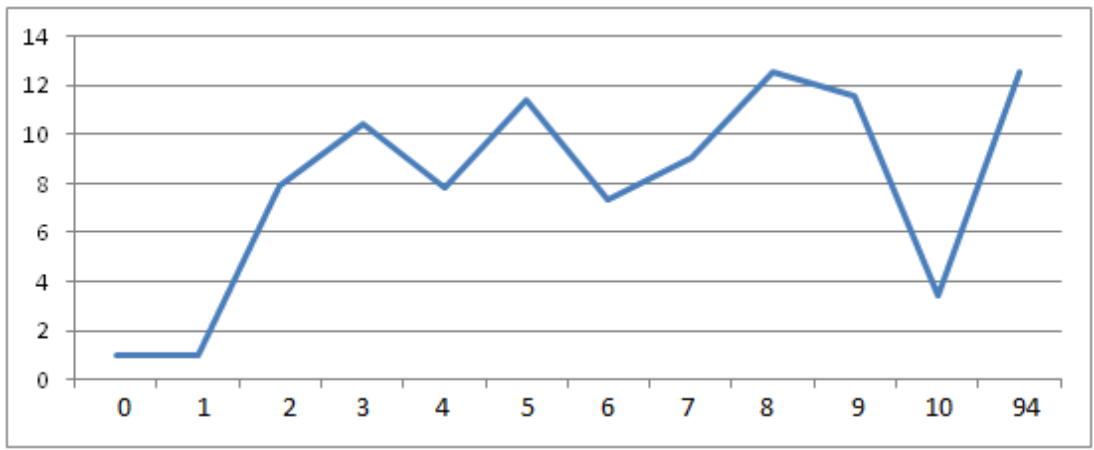

Figure 9 Improvement Factor of the d-Sequence of 1787 when divided into 94 blocks of size 19 digits each that is odd number of blocks

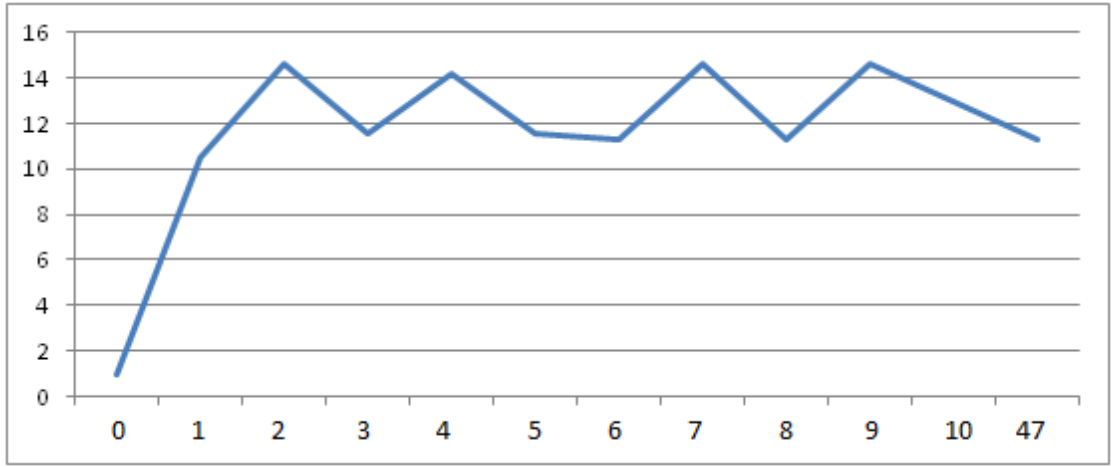

Figure 10 Improvement Factor of the d-Sequence of 1787 when divided into 47 blocks of size 38 digits each that is odd number of blocks

From all the above experiments it is found that the randomness of a d-sequence increases by applying permutations on its blocks. Similar results are obtained for a random sequence that is generated on a Windows PC. The above graphs show that the improvement factor is quite impressive if the block size is not too small. Several statistical tests of randomness [5] were performed on the sequences and the results were supportive of the conclusion that the sequences are cryptographically strong.

**Conclusion**

We show that permutations on blocks of random sequences improve their randomness. The improvement presented in the graphs is typical of the performance of d-sequences. The specific conclusion is that two or three permutations on blocks that are not too small suffice to improve the autocorrelation function of the sequence.

## References

[1]  C. H. Bennett and G. Brassard, "Quantum cryptography: Public key distribution and coin tossing," in Proceedings of the IEEE International Conference on Computers, Systems, and Signal Processing, Bangalore, India, pp. 175–179 (IEEE, New York, 1984).
[2]  S. Kak, A three-stage quantum cryptography protocol. Foundations of Physics Letters 19: 293-296, 2006.
[3]  Y. Chen, P. Verma, and S. Kak, Embedded security framework for integrated classical and quantum cryptography in optical burst switching networks. Security and Communication Networks 2: 546-554, 2009.

[4]  S. Kak, Quantum information and entropy, International Journal of Theoretical Physics 46: 860-876, 2007.
[5]  A. Rukhin *et al*, A Statistical Test Suite for Random and Pseudorandom Number Generators for Cryptographic Applications. NIST, 2010.
[6]  A.N. Kolmogorov, Three approaches to the quantitative definition of information. Problems Inform. Transmission 1 (1): 1–7, 1965.
[7]  S. Kak, Active agents, intelligence, and quantum computing. Information Sciences 128: 1-17, 2000.
[8]  S. Kak, The universe, quantum physics, and consciousness. Journal of Cosmology 3: 500-510, 2009.
[9]  S. Golomb, Shift Register Sequences. San Francisco, Holden–Day, 1967
[10] S. Kak and A. Chatterjee, On decimal sequences. IEEE Transactions on Information Theory, IT-27: 647 – 652, 1981.
[11] S. Kak, Encryption and error-correction coding using D sequences. IEEE Transactions on Computers C-34: 803-809, 1985.
[12] S. Kak, New results on d-sequences. Electronics Letters 23: 617, 1987.
[13] D. Mandelbaum, On subsequences of arithmetic sequences. IEEE Trans on Computers 37: 1314-1315, 1988.
[14] S Kak, Prime reciprocal digit frequencies and the Euler zeta function. http://arxiv.org/ftp/arxiv/papers/0903/0903.3904.pdf
[15] S. Rangineni, Cryptographic hardening of d-sequences. http://arxiv.org/abs/1106.3574